\documentstyle[prl, aps, twocolumn, epsfig]{revtex}

\newcommand\be{\begin{equation}}
\newcommand\ee{\end{equation}}
\newcommand\bd{\begin{displaymath}}
\newcommand\ed{\end{displaymath}}
\newcommand\bea{\begin{eqnarray}}
\newcommand\eea{\end{eqnarray}}

\begin{document}

\twocolumn[

\hsize\textwidth\columnwidth\hsize\csname @twocolumnfalse\endcsname

\draft

\title{Patterns from Preheating}

\author{A. Sornborger}
\address{NASA/Fermilab Astrophysics Group, Fermi National Accelerator
Laboratory, Box 500, Batavia, IL 60510-0500, USA}
\author{M. Parry}
\address{Department of Physics, Brown University, Providence, RI 02912}
\date{\small November 19, 1998}

\maketitle

\begin{abstract}

The formation of regular patterns is a well-known phenomenon in
condensed matter physics. Systems that exhibit pattern formation are
typically driven and dissipative with pattern formation occurring in
the weakly non-linear regime and sometimes even in more strongly
non-linear regions of parameter space. In the early universe,
parametric resonance can drive explosive particle production called
preheating. The fields that are populated then decay quantum
mechanically if their particles are unstable. Thus, during preheating,
a driven-dissipative system exists. In this paper, we show that a
self-coupled inflaton oscillating in its potential at the end of
inflation can exhibit pattern formation.
\\ \\Fermilab Preprint: Pub-98/373-A, BROWN-HET-1153, November 1998
\end{abstract}


]

Much recent work has been done on the topic of preheating in
inflationary cosmology. Preheating is a stage of explosive particle
production which results from the resonant driving of particle modes
by an inflaton oscillating in its potential at the end of inflation
\cite{trasbran,kls94,stb95}.

In regions of parameter space where parametric resonance is effective,
much of the energy of the inflaton is transferred to bands of resonant
wave modes. This energy transfer is non-thermal and can lead to
interesting non-equilibrium behavior. Two examples of the non-equilibrium
effects that can be produced are non-thermal phase transitions
\cite{kofmlind,tkac,rt96,khlekofm1} and baryogenesis
\cite{andelind,kolblind}. The non-thermal phase transitions induced during
preheating can also possibly lead to topological defect formation
\cite{kasukawa,tkackhle,parrsorn}, even at energies above the eventual
final thermal temperature. Furthermore, non-linear evolution of the field
when quantum decay of the resonantly produced particles is small leads to
a turbulent power-law spectrum of density fluctuations
\cite{khletkac,pr97}. 

In this letter, we present a new phenomenon that can arise from
preheating: pattern formation. It has long been known that many condensed
matter systems exhibit pattern formation\footnote{For an extensive review
of pattern formation in condensed matter systems, see
\cite{croshohe,newepass}}. Examples of pattern forming systems which have
been studied are ripples on sand dunes, cloud streets and a variety of
other convective systems, chemical reaction-diffusion systems, stellar
atmospheres and vibrated granular materials. All of these physical systems
have two features in common. They are all driven in some manner, i.e.
energy is input to the system, and they are all dissipative, usually being
governed by diffusive equations of motion. Typically, patterns are formed
in these systems in the weakly non-linear regime before the energy
introduced into the system overwhelms the dissipative mechanism.
Sometimes, patterns persist beyond the weakly non-linear regime as well. 

At the end of inflation, the inflaton $\phi$ is homogeneous with small
perturbations $\delta \phi$ imprinted on it due to quantum fluctuations. 
In chaotic inflationary models, the inflaton then oscillates about the
minimum of its potential, giving an effectively time dependent mass to
fields with which it is coupled. The time dependent mass drives
exponential growth in the population of bands of particle wave modes. Many
of the fields into which the inflaton can decay resonantly are also
unstable to quantum decay. For these reasons, at the end of inflation, we
are considering fields which are driven, due to resonant particle
creation, and also dissipative, due to quantum decay. Therefore, it makes
sense to see if there is a region of parameter space where the system is
in the weakly non-linear regime and exhibits pattern formation.

In our investigation we consider $\lambda \phi^4$ theory with the addition
of a phenomenological decay term to mimic the inflaton's quantum decay. 
This model without the decay term has been studied extensively in the
literature\cite{kls94,stb95,khletkac,bvhs96,pr97,k97,greekofm}, and a
similar model including the decay term has also been studied
\cite{kolbriot}. 

Our field equation is
\begin{equation}
  \ddot\phi + \gamma\dot\phi - \nabla^2\phi + \lambda\phi^3 = 0
\end{equation}
where $\gamma$ is a decay constant and $\lambda$ is the self-coupling
of the field. Here, we neglect the expansion of the universe, although
we will comment on the effect of expansion later. For our
calculations, we rescale: $t \rightarrow t/\sqrt{\lambda}\phi_0$, $x
\rightarrow x/\sqrt{\lambda}\phi_0$ and $\phi \rightarrow \phi\,\phi_0$,
where $\phi_0$ is the value of the inflaton at the
end
of inflation. This gives us a new equation \begin{equation}
  \ddot\phi + \Gamma\dot\phi - \nabla^2\phi +
    \phi^3 = 0 
\end{equation}
where $\Gamma = \gamma/\sqrt{\lambda}\phi_0$.

It should be noted that pattern formation in the inflaton system is
conceptually distinct from condensed matter systems for at least two
reasons. First, the equations we study are wave equations with
damping, not diffusive equations. Secondly, we expect wave patterns to
be formed while the homogeneous mode decays, therefore pattern
formation will be a temporary phenomenon, at least in the model above
in which gravity is neglected. This should be considered in contrast
to the typical condensed matter system, in which energy is introduced
via boundary conditions (in a convective system) or by a vibrating bed
(in a granular material system), and the energy input is essentially
constant.

For $\Gamma = 0$, the resonant modes lie in the interval
\begin{equation}
  \frac{3}{2} < k^2 < \sqrt{3}.
\end{equation}

For $\Gamma \neq 0$, we can introduce $\varphi = \phi\,
e^{\frac{\Gamma}{2} t}$ giving
\begin{equation}
  \ddot\varphi - (\nabla^2 + \frac{\Gamma^2}{4})\varphi 
    + e^{-\Gamma t}\varphi^3 = 0.
\end{equation}
Therefore, for small $\Gamma$, we expect the resonance bands to be
slightly shifted, because the $\nabla^2$ term is shifted by $\Gamma^2/4$,
and we expect the resonance to diminish slowly over time, due to the
exponential damping of the potential with time. It should be noted however
that the resonance structure of the equation can be quite sensitive to
changes in the potential\cite{greekofm}. 

As the effect that we are trying to isolate is non-linear, we
resort to numerical simulation of the field equation. We use a
leapfrog code which is second-order accurate in time and we use
fourth-order spatial differences. We simulate the field in 
two dimensions in a box with periodic boundary conditions which has
$256$ grid points per dimension.

For initial conditions, we give small amplitudes ($\sim 10^{-3}$) to
all resonant modes in the box. It is unnecessary to populate
non-resonant modes because they will decay while the resonant modes
grow. In the initial condition setup, it is crucial that there be many
resonant modes in the box.

Early on, we found spurious pattern formation when we set the box size
such that the resonant mode wave number in the box was $3$. For such small
wave numbers, only two resonant modes exist in the box, one along the
$x$-axis, and one along the $y$-axis, giving rise to a misleading square
wave pattern. Setting the box size such that the resonant wave number in
the box is $16$ is enough to give many different resonant modes in the
box, but still have good resolution of the wave, so this is the box size
we used. 

We set out to identify the weakly non-linear regime. In \cite{khletkac} it
is shown that the self-coupled inflaton system's non-linear time evolution
proceeds as follows: First, the resonant band amplitude grows. Next, when
the amplitude in the resonant band is high enough for non-linear effects
to become important, period doubling occurs and subsidiary peaks develop
in the power spectrum. Further peaks then develop and the spectrum
broadens and approaches a power-law spectrum. We tuned $\Gamma$ such that
the amplitude of the resonant band grew, but little period doubling
occurred. In this regime, only resonant mode wavelengths would exist in
the box and they would interact with each other non-linearly. 

\begin{figure}
\psfig{figure=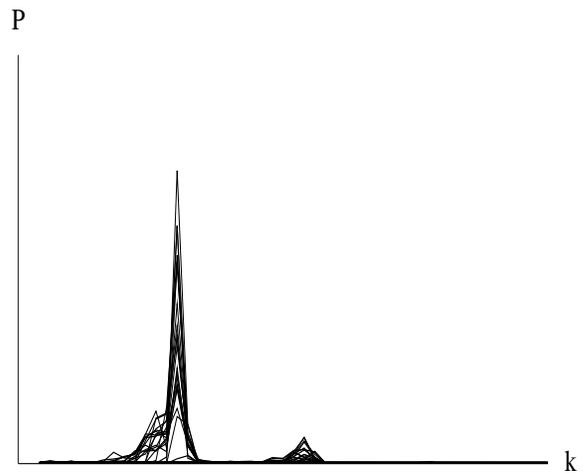,height=2.5in,width=3.0in}
\caption[caption]{A superposition of power spectra taken from a
simulation with $\Gamma = 0.0035$. We plot the amplitude of the power
spectra vs. wave number. Note that the period doubling modes are only
weakly populated, indicating that the simulation is in the weakly
non-linear regime.
\label{pattpower}}
\end{figure}

In figure \ref{pattpower} we plot a superposition of the power
spectrum at various times during our simulation. It is possible to see
that for the value $\Gamma = 0.0035$, the system stays in the weakly
non-linear regime for the entire simulation.

\begin{figure}
\psfig{figure=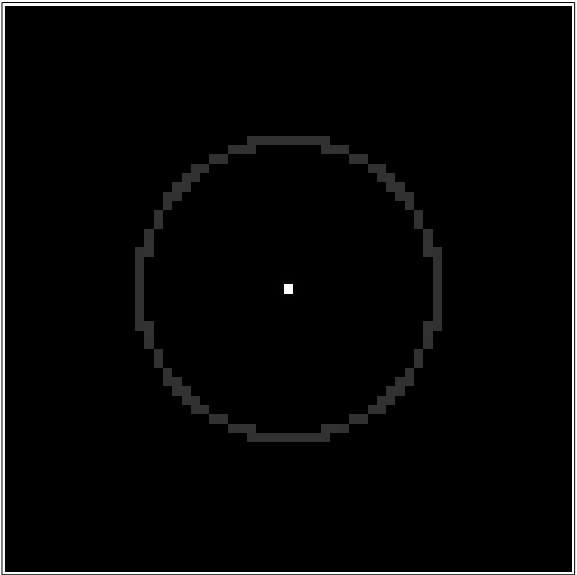,height=3.0in,width=3.0in}
\caption[caption]{Initial conditions for the
simulation ($\Gamma = 0.0035$). $\tilde\phi(k_x, k_y)$, the Fourier
transform of the inflaton is plotted. The central peak is the zero
mode, and the surrounding ring is the resonant mode populated with
small amplitudes. $t = 0$. Only the region of interest is plotted. All
modes outside of the plotted region have zero amplitude.
\label{pattk01}}
\end{figure}

\begin{figure}
\psfig{figure=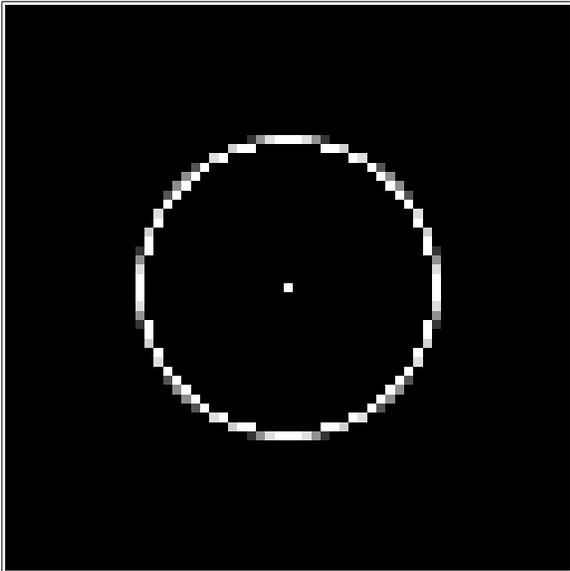,height=3.0in,width=3.0in}
\caption[caption]{After resonance begins to boost the amplitude of the
resonant mode. Notice the brightening (increasing amplitude) of the
resonant modes. $t = 60$.
\label{pattk02}}
\end{figure}

\begin{figure}
\psfig{figure=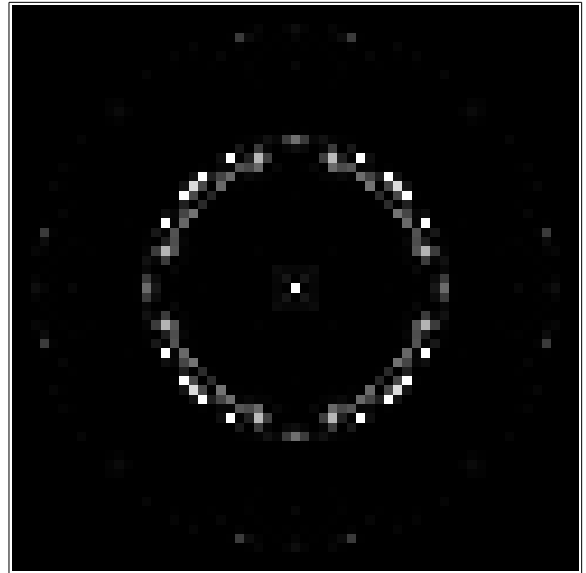,height=3.0in,width=3.0in}
\caption[caption]{Non-linear effects have become important by this
time causing fragmenting of the resonant mode. Modes with dominant
growth are taking over. $t = 210$.
\label{pattk07}}
\end{figure}

\begin{figure}
\psfig{figure=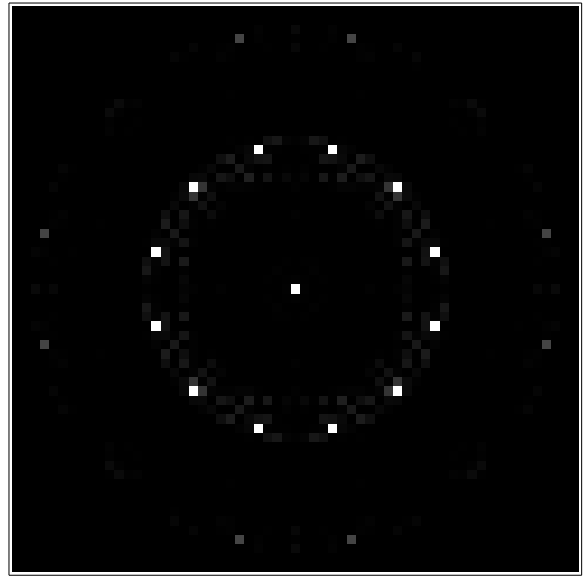,height=3.0in,width=3.0in}
\caption[caption]{The final wave pattern. $t = 300$.
\label{pattk10}}
\end{figure}

\begin{figure}
\psfig{figure=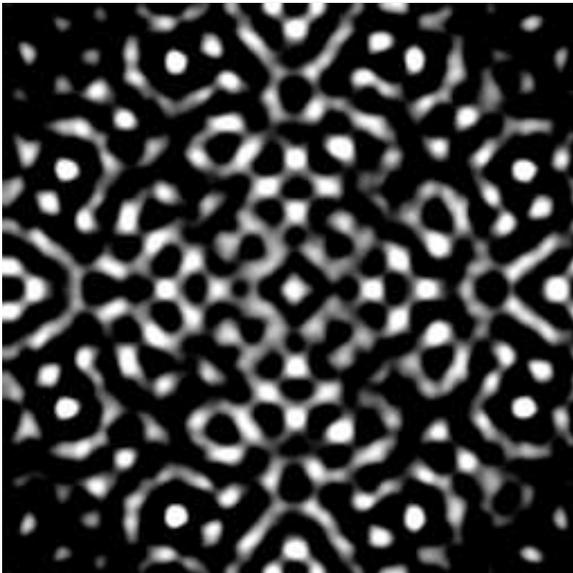,height=3.0in,width=3.0in}
\caption[caption]{The final wave pattern in configuration space. $t =
300$. The corners are darkened by small amplitude long wavelength
modes which modulate the wave pattern.
\label{pattout10}}
\end{figure}

When wave patterns form, the specific pattern which arises is due to the
non-linear interaction of the wave modes. The amplitude of wave modes
separated by different angles grows at different rates. Modes separated by
angles with the fastest growing amplitudes dominate the solution and form
the wave pattern. In plots \ref{pattk01}, \ref{pattk02}, \ref{pattk07} and
\ref{pattk10}, we show the evolution of the Fourier transform of the
inflaton. In the final plot, it is clear that the dominant modes have
picked out a preferred configuration in Fourier space. In figure
\ref{pattout10}, we also plot $\phi(x, y)$ in configuration space at the
time when the pattern has formed. It should be commented that the
preferred angles in Fourier space of the pattern would pick out a
dodecahedron, but this is not a close-packable structure, therefore the
pattern looks more complicated in configuration space, than in Fourier
space. After this time, the field amplitude begins to decay; the wave
pattern remains imprinted, but decreases in amplitude. 

The important result here is not that we have found the preferred
wave pattern of the system. With only a discrete set of wavemodes in
the box, we cannot be sure that some other pattern would not dominate
if we were able to investigate a continuum of modes. However, it is
clear that some pattern will be picked out by the system in this
parameter region. Therefore, we can claim to have identified pattern
formation as a phenomenon exhibited by the damped, self-coupled
inflaton.

It would also seem that pattern formation should exist in realizations
of parametric resonance and preheating other than the system we have
analyzed, simply because of their driven-dissipative nature.

In an expanding universe, after a similar variable and field
transformation as before,
the field equations (in conformal time) become
\begin{equation}
  \ddot\varphi + a\Gamma\dot\varphi - \nabla^2\varphi 
    - (\dot a\Gamma + \frac{\ddot a}{a})\,\varphi + \varphi^3 = 0,
\end{equation}
where here $\varphi = a\phi$. Note that there is no Hubble damping term
because the theory without damping due to quantum decay is conformally
invariant. (We are studying other models though, to see whether Hubble
damping can provide the necessary damping for pattern formation.) For the
expanding case, we expect pattern formation for smaller values of $\Gamma$
as the effective dissipation grows with the scale factor $a$. Oscillation
of the homogeneous mode should be dominated by the $\varphi^3$ term at
early times before $\varphi$ has had time to be significantly damped. We
are currently investigating this case.

\noindent

For wave patterns to have an impact cosmologically, they must persist
after they are created. In our toy model, the resonance band corresponds
to wavelengths roughly the size of the Hubble radius at the start of
preheating, about $10^{-25}\,cm$. The size of these fluctuations today
would then only be about $10\,m$. However we have seen that the patterns
are modulated by long wavelength modes, and it is this non-linear transfer
of power to small $k$-modes that may be important. Thus the size of the
Hubble radius at the {\it end} of preheating may actually set the scale
for wave patterns. In any case, as the wave patterns are regions of energy
density and are present essentially from the beginning of the universe,
they may be sites of significant gravitational accretion. As dark matter
seeds it may be possible for wave patterns to change the thermal history
of baryonic matter. It is this direction we intend to take in determining
the role wave patterns may play in the history of the universe. 

\vskip0.5cm

{}\noindent{\bf Acknowledgements}

It is a pleasure to be able to thank Rocky Kolb and Robert Brandenberger
for a number of useful discussions.  This work was supported by the DOE
and the NASA grant NAG 5-7092 at Fermilab. A portion of the computational
work in support of this research was performed at the Theoretical Physics
Computing Facility at Brown University.

\end{document}